\documentstyle[prl, twocolumn, aps, psfig]{revtex}
\begin{document}
\draft
\preprint{UCRL-JC-   }
\title{Binary Neutron-Star Systems: From the Newtonian Regime to the Last
Stable Orbit}

\author{P. Marronetti$^1$, G. J. Mathews$^1$ and J. R. Wilson$^2$}

\address{
$^1$University of Notre Dame,
Department of Physics,
Notre Dame, IN 46556}

\address{
$^2$University of California,
Lawrence Livermore National Laboratory,
Livermore, CA  94550}

\date{\today}
\maketitle
\begin{abstract}
We report on the first calculations of fully relativistic binary
circular orbits to span a range of separation distances 
from the innermost stable circular orbit (ISCO), deeply inside 
the strong field regime, to a distance ($\sim$ 200 km) where 
the system is accurately described by Newtonian dynamics.
We consider a binary system composed of two identical 
corotating neutron stars, with 1.43 $M_\odot$ gravitational mass 
each in isolation.
Using a conformally flat spatial metric we find solutions
to the initial value equations that correspond to semi-stable
circular orbits. At large distance, our numerical results agree 
exceedingly well with the Newtonian limit.
We also present a self consistent determination
of the ISCO for different stellar masses.
\end{abstract}

\narrowtext
Neutron-star binary systems are currently of interest as
sources of gravitational radiation \cite{gwaves,c93}.
The development of a new generation of laser interferometric
and cryogenic gravitational wave detectors
\cite{detectors} has renewed interest in theoretical models 
for generating gravitational radiation.
Binary systems composed by neutron stars 
and/or black holes are the most promising sources of detectable
gravitational radiation. However, the expected signal-to-noise
ratio is so low that the extraction of information will be difficult
without good theoretical waveforms.
Thus, it is important to accurately model such systems.

At the frequencies of interest for laser interferometric (10-100 Hz)
and cryogenic ($\sim$ 1000 Hz) detectors the orbits 
enter the strong gravity regime. Expansion methods applicable in the
Newtonian limit begin to breakdown before the strong field regime.
Hence, it becomes imperative to model 
the binary evolution with a method which accurately extends from
the Newtonian limit to the strong gravity regime.
In this paper we present the first such calculation. These
calculations are for one particular scenario, namely that of two
equal-mass stars constrained to rigid corotation in circular orbits.
Nevertheless, this represents a plausible benchmark for the
expected gravity wave signal.

Many groups have performed numerical simulations using different
approximations to this problem. Preliminary results have
been achieved using Newtonian dynamics \cite{Newtonian} or 
post-Newtonian expansion techniques \cite{PPN}. While these 
approximations work very well when the stars are 
very far from each other, they become unreliable when
the distance between stars reduces to a few stellar
radii.

Recently, a new fully relativistic approximation has been
used \cite{WM95,WMM96,MPW98}
to model binary neutron-star systems. This approximation is mainly
based upon a restriction that the spatial part of the metric tensor
is forced to be conformally flat (CFC). 
Work by Wilson and Mathews \cite{WM95} 
and Wilson, Mathews and Marronetti \cite{WMM96}
reported the first fully relativistic calculations of quasi-stable
circular orbits employing this approximation. The most
controversial \cite{collapse} of their results predicts that, 
for certain conditions, the stars could collapse into black 
holes prior to the merger. This controversy is irrelevant
to the present work in that it has been demonstrated
\cite{MPW98,Baumgarte} that this effect does not occur 
when rigid corotation is imposed.

Baumgarte et al. \cite{Baumgarte} recently developed a 
method for using the CFC to compute rigidly corotating stars. 
They found circular orbits for very close separation distances
(a few stellar radii) and estimated the point of secular 
instability of the stars. 
In the present work however, we implement a scheme to directly
determine the location of the ISCO for systems with different
masses. We also present numerical solutions for
quasi-stable circular orbits for corotating 
neutron stars that cover a wide range of separation distances between
the stars. Hence for the first time we present the much needed
calculation connecting the strong field regime at the 
{\it innermost stable circular orbit} (ISCO) with the weak field
region where the system is well described by Newtonian dynamics.

\vskip .1 in
A full discussion of the CFC method can be found in \cite{WMM96}. 
We use the (3+1) spacetime slicing as defined in
the ADM formalism \cite{adm62,y79}.
Utilizing Cartesian $x, y, z$ isotropic  coordinates, 
proper distance is expressed
\begin{equation}
ds^2 = -(\alpha^2 - \beta_n\beta^n) dt^2 + 2 \beta_n dx^n dt 
+ \gamma_{ns}dx^n dx^s~~,
\end{equation}
where the lapse function $\alpha$ 
describes the differential lapse of proper time between two
hypersurfaces.  
The quantity  $\beta_i$ is the shift vector denoting the shift in space-like
coordinates between hypersurfaces and $\gamma_{ij}$ is the spatial 
three-metric. The Latin indices go from 1 to 3.

Using York's (3+1) formalism \cite{y79},
the initial value equations can be written as follows;
the Hamiltonian constraint equation 
can be written,
\begin{equation}
R = 16\pi \rho + K_{ns}K^{ns} - K^2~~,
\label{ham}
\end{equation}
where $R$ is the Ricci scalar curvature, $K_{ns}$ is the extrinsic
curvature, and $\rho$ is the mass-energy density.

The momentum constraints have the form \cite{ev85}
\begin{equation}
D_n(K^{ni} - \gamma^{ni}K) = 8 \pi S^i~~,
\label{mom}
\end{equation}
where $D_n$ is the three-space covariant derivative and
$S^i$ are the spatial components of the four-momentum density.

The CFC method restricts the spatial metric $\gamma_{ij}$ to the form
\begin{equation}
\gamma_{ij} = \phi^4  \delta_{ij} ~,
\label{conftensor}
\end{equation}
where the conformal factor $\phi$ is a positive scalar function. This
approximation simplifies greatly the equations. It is motivated in 
part by the fact that the gravitational radiation in most systems
studied so far is small compared to the total gravitational
mass. The CFC is a frequently employed approach to the initial
value problem in numerical relativity. Its application here is
consistent with the quasi-equilibrium orbit approximation.
Further justification for its use can be found in refs. \cite{MPW98}
and \cite{Cook}.

The CFC leads to a set of elliptic 
equations for the metric components. Using Eq.~(\ref{ham})
in combination with the maximal slicing condition 
$tr(K)=0$, we get the following equations for $\phi$ and ($\alpha \phi$)
\begin{equation}
\nabla^2{\phi} = -4\pi\rho_1,
\label{phi}
\end{equation}
\begin{equation}
\nabla^2(\alpha\phi) = 4\pi\rho_2~~,
\label{alpha}
\end{equation}
where the $\nabla_i$ represent flat-space derivatives
and the source terms are

\begin{eqnarray}
\rho_1 =&& {\phi^5 \over 2}\biggl[\rho_{0} W^2 + 
\rho_{0}\epsilon ( \Gamma (W^2-1) + 1 )
+ {1 \over 16\pi} K_{ns}K^{ns}\biggr]
\label{rho1}
\end{eqnarray}
\begin{eqnarray}
\rho_2 = &&{\alpha \phi^5 \over 2}\biggl[\rho_{0} (3W^2-2)+
\rho_{0} \epsilon [ 3\Gamma (W^2+1)-5]\nonumber\\
&& + {7  \over 16\pi} K_{ns}K^{ns}\biggr]~~,
\label{rho2}
\end{eqnarray}
where $\rho_{0}$ is the rest-mass density,
$\epsilon$ the internal energy per unit of rest mass,
$\Gamma$ the adiabatic index, and $W$ a generalization of 
the special relativistic $\gamma$ factor \cite{WMM96}.
A solution of equation (\ref{alpha}) determines the lapse function after 
equation (\ref{phi}) is used to determine the conformal factor.

The shift vector $\beta^i$ can be decomposed \cite{Bow80}:

\begin{equation}
\beta^i = B^i - {1 \over 4}\nabla^i \chi~~.
\end{equation}
This is introduced into Eq.~(\ref{mom}) to obtain
the following two elliptic equations

\begin{equation}
\nabla^2 \chi    = \nabla_n B^n~~,
\label{chibeta}
\end{equation}
\begin{equation}
\nabla^2 B^i    = 2 \nabla_n ln(\alpha \phi^{-6}) K^{in} 
- 16 \pi \alpha \phi^4 S^i ~.
\label{capb}
\end{equation}

An equation for the extrinsic curvature $\hat K^{ij}$ is derived
\cite{WMM96} using the time evolution equation and the maximal
slicing condition

\begin{equation}
\hat K^{ij} = {\phi^6\over 2 \alpha} (\nabla^i \beta^j+
\nabla^j \beta^i - {2\over 3} \delta_{ij} \nabla_n \beta^n)~~,
\end{equation}
where $\hat K^{ij}=\phi^{10} K^{ij}$.

We assume that the matter behaves like a perfect fluid 
with a stress-energy tensor

\begin{equation}
T^{\mu\nu}=(\rho_{0}(1+\epsilon)+P) u^\mu u^\nu + P g^{\mu\nu}~~,
\label{stress-energy}
\end{equation}
 and use a polytropic equation of state (EOS)
\begin{equation}
P = k ~\rho_{0}^{\Gamma}~~,
\end{equation}
with $P$ the pressure and $k$ a constant.
The results presented here are for stars with $\Gamma=2$.

Following Baumgarte et al. \cite{Baumgarte},
for rigidly corotating stars, the fluid four velocity 
can be taken as proportional proportional to a Killing vector

\begin{equation}
{\partial \over\partial t} + \omega {\partial \over\partial \Phi}~~,
\end{equation}
where $\omega$ is the orbital angular frequency and $\Phi$ the azimuth
coordinate.
In this case the steady-state limit of the hydrodynamics 
momentum equation \cite{MPW98} yields the relativistic Bernoulli 
equation \cite{Baumgarte}

\begin{equation}
q = {1\over {1+n}} \biggl( { 1+C \over {\alpha
(1-v^2)^{1/2}}} - 1 \biggr)~~,
\label{Bernoulli}
\end{equation}
where $q=P/\rho_{0}$, $C$ is a constant of integration and $v$
is the matter proper velocity \cite{problems}.
From Eqs.~(\ref{stress-energy}) and (\ref{Bernoulli}) we find 
expressions for the proper baryonic matter density $\rho_{0}$ and the 
material momentum density $S^i$ as functions of the fields. We choose 
$\alpha$, $\beta^i$, $\phi$, and $q$ as the independent 
variables of our set of equations. 

\vskip .1 in

The set of equations is solved numerically using an iterative
algorithm based upon a specially designed elliptic solver.
This method consists of a combination of multigrid algorithms
and domain decomposition techniques \cite{MM98}. 
The three dimensional spatial volume where the equations
are solved is divided into concentric layers. These layers are centered
around the star and the grid resolution decreases with the distance
from the stars. To avoid bias, each iteration cycle starts 
assuming zero angular frequency and spherical stars.
The equations are solved iteratively until numerical 
convergence for the fields and the frequency is achieved.
Our calculations utilize 40 grid points 
on average across the stellar diameter. This is about twice as
large as the number used in previous works \cite{Baumgarte}.
This efficient use of computer memory permits us to
describe adequately the interior of the stars even when 
they are so far apart that they approach the regime of 
Newtonian point masses. 

The boundary conditions are estimated from
the first terms in a multipole expansion of the fields.
This method is quite accurate when the boundary surfaces are 
very far from the stars, which is the case in our computations.
Thus, these methods allow us to connect between the strong 
gravitational field regime close to the ISCO and the Keplerian regime.
Domain decomposition techniques also provide a natural way
of code parallelization, which reduces the processing time enormously
\cite{MM98}.

Solutions are obtained for specific values of two free
parameters, namely, the coordinate distance between stars and
the stellar baryonic mass. The reason to choose the latter as a free
parameter is that the baryonic mass of the star 
remains constant during the inspiral that describes
most of the time evolution of the system.
This differs from the approach of \cite{Baumgarte} who used
central density and separation distances as free parameters and
interpolated to find the solutions of constant baryonic mass.
In our calculations we iterate to obtain a given baryonic mass.
This avoids the numerical error associated with interpolation.
Thus, we were able to construct constant baryonic mass 
sequences of orbits with a minimum number of code runs.

Using the CFC approximation we found solutions to the initial value
equations for semi-stable circular orbits for a binary system 
of identical neutron stars with 1.55 $M_\odot$ baryonic mass 
and 1.43 $M_\odot$ gravitational mass in isolation. 
The former values represent a typical neutron star and were 
obtained fixing by the value of the polytropic constant
(in polytropic models physical quantities can be rescaled with 
the polytropic constant \cite{Baumgarte}).

\begin{figure}[htb]
\begin{center}
\hskip 2.0 cm
\vskip 0.5 cm
\mbox{\psfig{figure=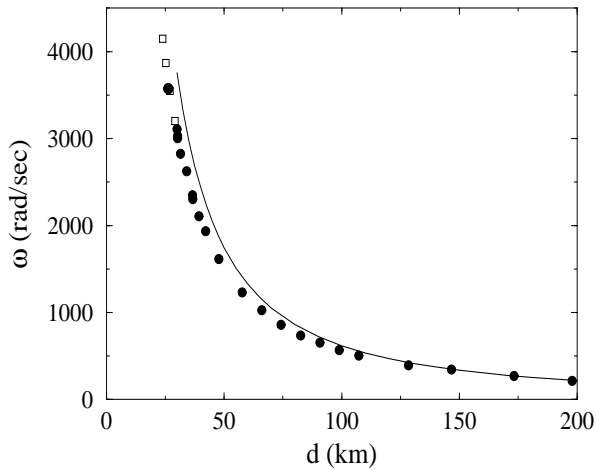,width=6 cm,height=5 cm}}
\end{center}
\vskip 0.7 cm
\caption{Orbital frequency as a function of the coordinate
separation between star centers. The solid line represents
Kepler's law, the circles are the orbits corresponding
to a sequence of stellar baryonic mass of 1.55 $M_\odot$,
and the squares are results from Baumgarte et al. [10].}
\label{ovsd}
\end{figure}

Figure \ref{ovsd} shows the angular frequency as seen by 
a distant observer as a function of the coordinate distance
between the centers of the stars (circles).
The coordinate distance is not a gauge invariant quantity,
but since it converges asymptotically
to flat space separations, it allows us to compare our results
with Kepler's law (solid line). We can also compare our results with those
from Baumgarte et al. \cite{Baumgarte} (squares) since 
we use the same coordinates. They estimate the onset of 
secular instability after which the stars can no longer 
maintain rigid corotation. They assume that the ISCO is close 
to this point since the secular instability is expected to occur 
just before  the dynamical instability that defines the ISCO \cite{Lai93}.
Note that their calculations include numerical solutions for
orbits that are inside the ISCO which are used to determine
the turning point in the binding energy \cite{Baumgarte}.

\begin{figure}[htb]
\begin{center}
\hskip 2.0 cm
\vskip 0.5 cm
\mbox{\psfig{figure=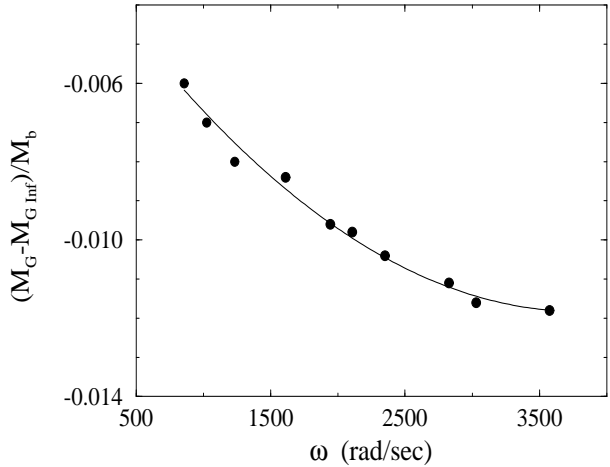,width=6 cm,height=5 cm}}
\end{center}
\vskip 0.7 cm
\caption{Binding energy of the system as a function of the orbital
frequency as seen by a distant observer. The points represent the
orbits of the high resolution runs and the solid line corresponds
to a polynomial fit. We identify the lowest point as the ISCO of 
the system.}
\label{BEvso}
\end{figure}

The numerical method developed for this work shows the 
absence of solutions for circular orbits
when the stars are too close.
This situation is consistent with the existence of a dynamical
instability which defines an innermost stable circular orbit
for the binary system.
However, there is still the possibility that this absence
is due to a problem in the convergence of the numerical scheme.
The ISCO determined here is thus the stable orbit with the 
highest angular frequency (in the sequence shown in Fig. \ref{ovsd},
it is at $\omega=3576$ rad/sec).

The value of angular momentum at the ISCO is $J=1.61\times 10^{11}~cm^2$,
which gives a value of $J/{M_G}^2=0.94$. Thus, the stars could
merge into a Kerr black hole without the further loss of angular
momentum.

Figure \ref{BEvso} shows the binding energy as a function of the 
angular frequency of the system as seen by a distant observer.
The binding energy is represented as one half the ADM mass of the
binary minus the gravitational energy of an isolated star, divided
by the baryonic mass of the isolated star. The points correspond
to the orbits obtained using high resolution (more than 30 grid
points across the stellar diameter) and the solid line corresponds
to a polynomial fit.
The lowest plotted value of the binding energy on Fig.~\ref{BEvso} 
coincides with the orbit we identify as the ISCO. In this calculation
no turning point was observed (within the numerical error). 
For large separations (low angular frequencies) the value of the 
binding energy approaches zero as expected.

\begin{figure}[htb]
\begin{center}
\hskip 2.0 cm
\vskip 0.5 cm
\mbox{\psfig{figure=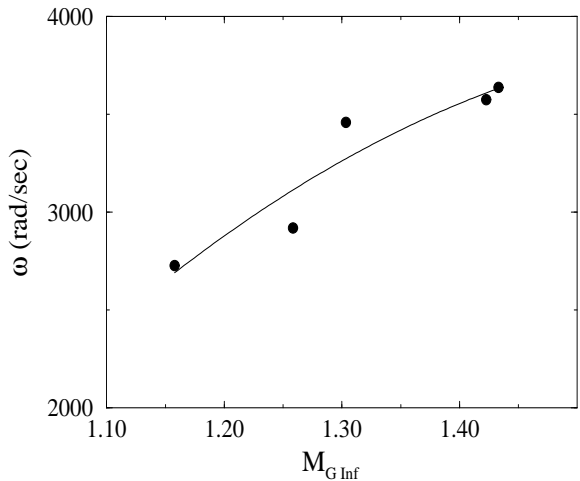,width=6 cm,height=5 cm}}
\end{center}
\vskip 0.7 cm
\caption{Orbital angular frequency of the ISCO for stars with different
gravitational masses (in units of solar masses) in isolation.}
\label{ISCO}
\end{figure}

Figure \ref{ISCO} shows the ISCO angular frequency for systems of stars
with different gravitational mass in isolation. 
The angular frequencies shown in Fig.~\ref{ISCO} are somewhat 
lower ($\sim 10 \% $) than the estimations corresponding to the secular 
instability points from Baumgarte et al. \cite{Baumgarte}.
This is most likely due to effects of grid resolution.

The excellent agreement between our results at large separations
and the values obtained from Newtonian dynamics 
provides an additional check on the numerical code.

We observe a slight decrease in the central density when the stars
approach each other similar to the results reported in 
\cite{Baumgarte}. This is contrary to what is observed in 
fully hydrodynamical simulations. In reference \cite{MPW98} 
however it has been shown analytically and numerically that no 
increase in central density occurs for stars in 
rigid corotation. This also confirms that the effect is not 
an artifact of the CFC method.

An analysis of the emission of gravitational radiation along
sequences of constant baryonic mass will be the subject of a forthcoming
article.

\vskip .1 in

Work at University of Notre Dame supported in part by
NSF grant PHY-97-22086. Work performed in part under the auspices 
of the U.~S.~Department of Energy
by the Lawrence Livermore National Laboratory under contract
W-7405-ENG-48.


\begin{references}

\bibitem{gwaves}J. P. A. Clark and D. M. Eardley, Astrophys. J., {\bf 215}, 311 (1977);
J. P. A. Clark E. P. J. van den Heuvel, and W. Sutantyo, 
Astron. \& Astrophys., {\bf 72}, 120 (1979); K. Thorne, in 
{\it 300 Years of Gravitation}, S. Hawking and W. Israel,
eds., (Cambridge: Cambridge Univ. Press), 378 (1987);
B. F. Schutz, Nature, {\bf 323}, 310 (1986);
B. F. Schutz, Classical Quantum Gravity, {\bf 6}, 1761 (1989);

\bibitem{c93}C. Cutler, {\it et al.}, Phys Rev. Lett., {\bf 70}, 2984 (1993);
F. A. Rasio and S. L. Shapiro, Astrophys. J., {\bf 432}, 242 (1994); K. Oohara and
T. Nakamura,  Prog. Theo. Phys., {\bf 88}, 307 (1992).

\bibitem{detectors}E. Amaldi, {\it et al.}, Astron. Astrophys., 
{\bf 216}, 325 (1989);
A. Abramovici, {\it et al.}, Science, {\bf 256}, 325 (1992);
C. Bradaschia, {\it et al.}, Nucl. Instrum. Meth., Phys. Res. Sect., {\bf 289}, 
518 (1990).

\bibitem{Newtonian}F. A. Rasio and S. L. Shapiro, Astrophys. J. {\bf 401},
226 (1992); {\bf 432}, 242 (1994); M. Shibata, T. Nakamura and K.
Oohara, Prog. Theor. Phys. {\bf 88}, 1079 (1992); X. Zhuge, J. M.
Centrella and S. L. W. McMillan, Phys. Rev. D {\bf 50}, 6247 (1994);
{\bf 54}, 7261 (1996); M. Ruffert, H-T. Janka and G. Schafer,
Astrophys. Sp. Sci. {\bf 231}, 423 (1993)

\bibitem{PPN}M. Shibata, Prog. Theor. Phys. {\bf 96}, 317 (1996); Phys.
Rev. D {\bf 55}, 6019 (1997); K. Taniguchi and M. Shibata, Phys.Rev. D
{\bf 56}, 798 (1997); M. Shibata and K. Taniguchi, Phys.Rev. D
{\bf 56}, 811 (1997); 

\bibitem{WM95}J. R. Wilson and G. J. Mathews, Phys. Rev. Lett. {\bf 75},
4161 (1995)

\bibitem{WMM96}J. R. Wilson, G. J. Mathews and P. Marronetti Phys. Rev.
D {\bf 54}, 1317 (1996)

\bibitem{MPW98} G. J. Mathews, P. Marronetti, and J. R. Wilson,
Phys. Rev. D. in press (gr-qc/9710140).

\bibitem{collapse} See Mathews, Marronetti and Wilson \cite{MPW98} 
and Shapiro \cite{Shapiro} for arguments on
why this effect was not seen by some previous works.

\bibitem{Baumgarte} T. W. Baumgarte, G. B. Cook, M. A. Scheel, S. L.
Shapiro and S. A. Teukolsky, Phys. Rev. Lett. {\bf 79}, 1182 (1997);
Phys. Rev. D {\bf57}, 6181 (1997).

\bibitem{adm62}R. Arnowitt, S. Deser, and C. W. Misner, in {\it Gravitation},
ed. L. Witten (New York: Wiley), p. 227 (1962).

\bibitem{y79}J. W. York, Jr., in {\it Sources of Gravitational 
Radiation}, ed. L . Smarr (Cambridge; Cambridge Univ. Press) p. 83 (1979).

\bibitem{ev85}C. R. Evans, PhD. Thesis, Univ. Texas (1985);
P. Anninos, D. Hobill, E. Seidel, L. Smarr, Phys Rev. Lett., {\bf 71},
2851 (1993).

\bibitem{Cook} G. B. Cook, S. L. Shapiro and S. A. Teukolsky, Phys. Rev.
D {\bf 50}, 2364 (1994)

\bibitem{Bow80} J. M. Bowen and J. W. York, Jr., Phys. Rev. D.
{\bf 21}, 2047 (1980).

\bibitem{problems} A. P. Lightman, W. H. Press, R. H. Price and S.
A. Teukolsky, {\it Problem Book in Relativity and Gravitation}
(Princeton University Press, Princeton, New Jersey, 1975).

\bibitem{MM98} P. Marronetti and G. J. Mathews, submitted to Journal
Comp. Phys. (1998).

\bibitem{Lai93}D. Lai, F. A. Rasio and S. L. Shapiro, Astrophys. J.
Suppl. {\bf 88}, 205 (1993).

\bibitem{Shapiro} S. L. Shapiro, Phys. Rev. D {\bf 57}, 908 (1998)

\end{references}
\end{document}